\documentclass[conference]{IEEEtran}
\IEEEoverridecommandlockouts

\usepackage{cite}
\usepackage{amsmath,amssymb,amsfonts}
\usepackage{algorithmic}
\usepackage{graphicx}
\usepackage{textcomp}
\usepackage{xcolor}
\usepackage{csquotes}
\usepackage{comment}
\usepackage[detect-all]{siunitx}
\usepackage{booktabs}

\usepackage{hyperref}

\usepackage{cleveref}
\crefname{figure}{Fig.}{Figs.}
\crefname{section}{Sec.}{Secs.}
\crefname{equation}{Eq.}{Eqs.}
\crefname{table}{Table}{Tables}

\usepackage{multirow}

\def\BibTeX{{\rm B\kern-.05em{\sc i\kern-.025em b}\kern-.08em
    T\kern-.1667em\lower.7ex\hbox{E}\kern-.125emX}}

\usepackage[nolist]{acronym}

\usepackage{enumitem}

\definecolor{improvement}{RGB}{34,139,34} 
\definecolor{decline}{RGB}{220,20,60} 

\newcommand{\improve}{\textcolor{improvement}{\hfill $\uparrow$}}
\newcommand{\decline}{\textcolor{decline}{\hfill $\downarrow$}}
\newcommand{\neutral}{\hfill $\sim$}

\begin{document}

\begin{acronym}
    \acro{rl}[RL]{Reinforcement Learning}
    \acro{ppo}[PPO]{Proximal Policy Optimization}
    \acro{drl}[DRL]{Deep Reinforcement Learning}
    \acro{mdp}[MDP]{Markov Decision Process}
    \acro{mpc}[MPC]{Model Predictive Control}
    \acro{1dof}[1-DoF]{one-degree-of-freedom}
    \acro{lqr}[LQR]{Linear-quadratic regulator}
    \acro{lqi}[LQI]{Linear-quadratic-integral regulator}
    \acro{lti}[LTI]{Linear Time-Invariant}
    \acro{mdp}[MDP]{Markov Decision Process}
    \acro{sb3}[SB3]{Stable Baselines3}
    \acro{trpo}[TRPO]{Trust Region Policy Optimization}
    \acro{icps}[ICPS]{Industrial \ac{cps}}
    \acro{cps}[CPS]{Cyber-Physical System}
    \acro{pid}[PID]{Proportional-integral-derivative}
\end{acronym}

\title{The Crucial Role of Problem Formulation in Real-World Reinforcement Learning
\thanks{Financial support for this study was provided by the Christian Doppler Association (JRC ISIA), the corresponding WISS Co-project of Land Salzburg, the European Interreg Österreich-Bayern project BA0100172 AI4GREEN and by the Federal Ministry of Education and Research (BMBF) under the project iCARus.}
}

\author{
\IEEEauthorblockN{
Georg Schäfer\IEEEauthorrefmark{1}\IEEEauthorrefmark{2}\IEEEauthorrefmark{3}, 
Tatjana Krau\IEEEauthorrefmark{4}, 
Jakob Rehrl\IEEEauthorrefmark{1}\IEEEauthorrefmark{2}, 
Stefan Huber\IEEEauthorrefmark{1}\IEEEauthorrefmark{2}, 
Simon Hirlaender\IEEEauthorrefmark{3}}
\IEEEauthorblockA{\IEEEauthorrefmark{1}Salzburg University of Applied Sciences, Salzburg, Austria}
\IEEEauthorblockA{\IEEEauthorrefmark{2}Josef Ressel Centre for Intelligent and Secure Industrial Automation, Salzburg, Austria}
\IEEEauthorblockA{\IEEEauthorrefmark{3}Paris Lodron University of Salzburg, Salzburg, Austria}
\IEEEauthorblockA{\IEEEauthorrefmark{4}University of Applied Sciences Kempten, Kempten, Germany \\
georg.schaefer@fh-salzburg.ac.at}
}

\maketitle

\begin{abstract}
Reinforcement Learning (RL) offers promising solutions for control tasks in industrial cyber-physical systems (ICPSs), yet its real-world adoption remains limited.
This paper demonstrates how seemingly small but well-designed modifications to the RL problem formulation can substantially improve performance, stability, and sample efficiency.
We identify and investigate key elements of RL problem formulation and show that these enhance both learning speed and final policy quality.
Our experiments use a one-degree-of-freedom (1-DoF) helicopter testbed, the Quanser Aero~2, which features non-linear dynamics representative of many industrial settings.
In simulation, the proposed problem design principles yield more reliable and efficient training, and we further validate these results by training the agent directly on physical hardware.
The encouraging real-world outcomes highlight the potential of RL for ICPS, especially when careful attention is paid to the design principles of problem formulation.
Overall, our study underscores the crucial role of thoughtful problem formulation in bridging the gap between RL research and the demands of real-world industrial systems.
\end{abstract}


\section{Introduction}

\ac{rl} is a machine learning paradigm where an agent learns to make decisions by interacting with an environment and iteratively adjusting its policy to maximize cumulative rewards~\cite{sutton2018reinforcement}.
This approach has shown impressive success in domains such as robotics~\cite{singh2022reinforcement}, video gaming~\cite{mnih2015human}, and \acp{cps}~\cite{liu2019reinforcement}.
In \ac{rl}, the agent does not require an a priori model of the environment; instead, it learns directly from data generated through interaction, making it particularly attractive for complex or uncertain scenarios.

\acp{icps} combine physical processes with computational control, forming the backbone of many modern industrial systems in manufacturing, energy management, and smart transportation~\cite{liu2019reinforcement}.
Conventional control approaches for \acp{icps} often rely on mathematical models, which are difficult to derive when facing dynamic, high-dimensional, or stochastic processes~\cite{sanchez2012simplied}.
\ac{rl} reduces this challenge by providing techniques that learn controllers or decision-making policies through direct experimentation or simulation. 
However, while \ac{rl} offers significant potential, its application in real-world \acp{icps} remains limited.
One main reasons for this is the lack of standardization in problem formulation and broader engineering workflows.

In comparison to supervised learning pipelines, where data preprocessing, model selection, training, evaluation, and deployment are well-defined steps~\cite{geron2022hands}, in \ac{rl} there are no universally systematic frameworks for problem design and experimentation established yet.
This gap frequently leads to an overemphasis on hyperparameter tuning at the expense of carefully specifying the environment, reward function, as well as state and action spaces.
Yet, thoughtful problem formulation is critical for ensuring that \ac{rl} solutions align with physical constraints (e.g., temperature limits, actuator torque bounds) and industrial objectives (e.g., safety, production quality).
In this paper, we claim that structured problem design is essential to closing the gap between \ac{rl} research and its real-world application in \acp{icps}.

\subsection{Contributions}
\begin{enumerate}
    \item We propose a set of structured problem-design principles for \ac{rl} in \acp{icps}.
    \item We systematically examine how factors such as normalization, target signal randomization, horizon lengths, initial state distributions, and action penalties influence training stability and performance.
    \item We validate these design principles on a \ac{1dof} helicopter testbed, highlighting their real-world applicability and efficiency.
    \item We demonstrate that \ac{rl} can be trained directly on physical hardware without relying on a priori models, underscoring its practical feasibility for industrial settings.
\end{enumerate}

\section{Design Principles of RL Problem Formulation} \label{sec:rl-problem-formulation}
Formulating the problem in \ac{rl} is crucial to the success of the learning process.
The interactions between the agent and the environment are often described by an infinite-horizon, discounted \ac{mdp} 
\[
M = (\mathcal{S}, \mathcal{A}, \mathcal{P}, r, \gamma, \mu)
\]

where $\mathcal{S}$ is the state space, $\mathcal{A}$ the action space, $\mathcal{P}$ the transition function, $r$ the reward function, $\gamma$ the discount factor, and $\mu$ the initial state distribution~\cite{agarwal2019reinforcement}.

However, in practical settings (particularly in real-world engineering tasks), one often adopts a \emph{finite-horizon} or \emph{episodic} approach to training and evaluation.
Although the horizon length~$T$ is not strictly part of the \ac{mdp} formalism, it plays a central role in how we structure each training episode, terminate interactions, and evaluate performance.
Following Puterman's classification~\cite{puterman2014markov}, the decision epochs may be bounded by a finite horizon or treated in an infinite-horizon framework.

\subsection{Components of Markov Decission Processes} 

\subsubsection{State Space}
The state space $\mathcal{S}$ defines all possible configurations the agent can encounter in the environment, forming the basis for decision-making.
It can be discrete or continuous, and must be designed to satisfy the Markov property, where future states depend only on the current state and action~\cite{sutton2018reinforcement}, influencing the choice of suitable \ac{rl} algorithms.
An effective state space must include all necessary information for optimal decision-making, while avoiding excessive complexity that leads to the \emph{curse of dimensionality}~\cite{kober2013reinforcement}.
When key information is missing from the state representation (partial observability), recurrent neural networks or additional sensor inputs may be required~\cite{xiang2021recent}.

\subsubsection{Action Space}
The action space $\mathcal{A}$ is the set of all possible actions the agent can take in any given state.
It may be discrete (e.g., specific commands) or continuous (e.g., real valued control adjustments).
An effective action space should reflect physical constraints of real-world systems.
Hierarchical action structures \cite{pateria2021hierarchical} for complex tasks may improve scalability and efficiency.
Details on action space shaping can be found in~\cite{kanervisto2020action}.

\subsubsection{Transition Function}
The transition function $P(s'|s,a)$ describes the environment's dynamics, determining the probability of transitioning from state $s$ to state $s'$ after taking action $a$.
The transition function should align with the Markov property, ensuring that future states depend solely on the current state and action.
This property is aligned with \ac{lti} state space models, where the following system state is determined solely by current inputs and the current state~\cite{chen1995linear}.
The shared principles of the Markov property in \ac{rl} and time invariance in control theory highlight the compatibility of \ac{rl} methods, such as those based on \acp{mdp}~\cite{kamthe2018data}, with control-theoretic frameworks.

In real-world applications, the transition function is generally not explicitly available, requiring either approximation through simulation or learning from interaction data.
Model-based \ac{rl} approximates or uses a known transition function to simulate dynamics, enabling greater sample efficiency by reducing the interactions with the environment~\cite{moerland2023model}.
However, inaccuracies in the model can lead to suboptimal policies.

In contrast, model-free approaches, such as Q-Learning \cite{watkins1992q} and PPO \cite{schulman2017proximal}, bypass this explicit modeling, relying on interaction data to learn policies or value functions directly, thus typically require more environment interactions to achieve comparable performance.
The choice between model-based and model-free methods depends on task complexity and computational constraints, with hybrid approaches emerging as promising solutions~\cite{chebotar2017combining}.

\subsubsection{Reward Function}
The reward function guides the agent by assigning rewards for transitions.
It is substantial to design the reward function with the task's goals.
Sparse rewards simplify the design, but provide limited feedback, which slows learning.
Dense rewards, on the other hand, accelerate learning by rewarding intermediate steps but risk introducing noise or instability if poorly designed.
Proper weighting of multiple objectives (e.g., efficiency and safety) is essential to avoid unintended behavior.
Reward shaping, a term formalized by Ng et al. in \cite{ng1999policy}, describes a technique in \ac{rl} where additional rewards are designed and added to the environment to guide the agent's learning process.
Balancing between sparse and dense rewards is a critical challenge that can significantly affect the efficiency and effectiveness of learning algorithms~\cite{eschmann2021reward, ibrahim2024comprehensive}.

\subsubsection{Discount Factor}
The discount factor $\gamma \in [0, 1)$ determines the importance of future rewards in the agent's decision-making process\cite{agarwal2019reinforcement}.
Higher $\gamma$ values encourage long-term planning, whereas lower $\gamma$ values prioritize immediate rewards~\cite{sutton2018reinforcement}.

\subsubsection{Initial State Distribution} 
The initial state distribution $\mu \in \Delta(\mathcal{S})$ where $\Delta(\mathcal{S})$ is the space of probability distributions over $\mathcal{S}$ defines the probability of the system starting in the initial state $s_0$~\cite{agarwal2019reinforcement}.
It plays a critical role in shaping the agent's initial interactions with the environment and the subsequent learning process.

\begin{itemize}
    \item \emph{Relevance:} The initial distribution should cover realistic scenarios to ensure the agent's learned policy is trained on the practical applications.
    \item \emph{Diversity:} A diverse initial state distribution can improve the agent's reliability by exposing it to a wide range of starting configurations, reducing the likelihood of overfitting to a specific set of initial conditions.
    \item \emph{Alignment:} The choice of $\mu$ should align with the task objectives and the environment's expected usage, ensuring that critical states are not underrepresented during training.
\end{itemize}

In practice, the initial state distribution can be defined deterministically (e.g., a fixed starting state) or stochastically (e.g., sampled from a predefined distribution).
Tasks involving exploration or environments with complex dynamics may benefit from stochastic initializations to avoid biasing the agent's learning toward specific regions of the state space.

\section{A Real-World RL Application} \label{sec:experimental-setup}

The testbed used for the experiments is the Quanser Aero~2 system, configured in its \ac{1dof} mode.
The system is equipped with two motors that control the pitch of the beam by adjusting the voltages applied to the motors.
We define the control task as achieving and maintaining desired pitch angles over time by applying appropriate voltage inputs.
The motors operate within a voltage range of \SIrange{-24}{24}{\volt}.
To determine the performance of the control strategy, the deviation to the target is considered.
Although relatively compact, the Quanser Aero~2 features non-linear system dynamics that mirror many of the challenges found in industrial control processes, making it an effective and representative platform for evaluation \ac{rl} approaches.

\subsection{Simulation and Real-World Environments}
The experiments were conducted in both a simulation environment and on the real-world environment.
A detailed explanation of how these environments were accessed can be found in~\cite{SSRHH24}.
A description of the model is proposed in~\cite{SRHH24}.

\begin{enumerate} 
    \item \emph{Simulation Environment:}
    The system was first modeled in Simulink to allow for a safe and controlled training process, meaning that no actual hardware is exposed to risk (e.g., the beam cannot crash into its socket or cause damage if a suboptimal policy is attempted).
    \item \emph{Real-world Environment:}
    The most promising problem formulation identified in simulation was then deployed directly on the physical Quanser Aero~2, to evaluate the feasibility of training on directly on the real hardware.
\end{enumerate}

\subsection{Performance Metrics}
The performance of the \ac{rl} agent in this study is primarily evaluated using the average deviation to the target.
This metric is defined as the mean absolute error between the actual pitch angle and the target pitch, and directly aligns with the base reward function used in the training process.
Other common control performance metrics, such as steady-state deviation, overshoot and rise-time (as proposed in \cite{SRHH24}), are excluded from this evaluation.

To evaluate the \ac{rl} training performance, three additional criteria are considered.
The first is training stability, measured by the standard deviation of returns across multiple training runs.
A lower standard deviation indicates more consistent performance, reflecting the reliability of the training process.
The second criterion is sample efficiency, measured as the number of training steps required to achieve the specified performance threshold of \ang{4} average deviation to the target.
This metric is particularly relevant for real-world applications where interaction time with the environment is limited.
Finally, the average absolute voltage applied to the system is measured, reflecting the energy efficiency of the policy.
All performance metrics were recorded on an evaluation environment identical to the environment from the baseline problem formulation.

\subsection{Baseline Problem Formulation}\label{sec:original-problem-formulation}

The original problem formulation in \cite{SRHH24} employed a time-limited setup with a fixed target profile.
This choice was motivated by practicality and by the goal of maintaining a straightforward, comparable setup.
Inspired by this formulation, we define the baseline problem formulation for our experiments as follows:

\begin{itemize} 
    \item \emph{State Space:} For each time step $t$, the state is defined as ($\Theta_{t}$, $\omega_t$, $r_t$), where $\Theta_{t}$ represents the pitch angle, $\omega_{t} = \Theta_{t}-\Theta_{t-1}$ denotes angular velocity, and $r_t$ is the desired target angle.
    This formulation was chosen to simplify the agent's state representation while still capturing the essential dynamics of the system, as the pitch and angular velocity naturally represent the states of the mechanical system.
    \item \emph{Action Space:} The action space consists of a single continuous action, which represents the voltage applied to the motors.
    This action space was bounded by the physical constraints of the system (\SIrange{-24}{24}{\volt}).
    The motors operate in opposition, meaning one motor applies positive voltage while the other applies an equal magnitude of negative voltage.
    \item \emph{Reward Signal:} The reward is defined as the negative absolute difference between the current pitch angle $\Theta_t$ and the target pitch angle $r_t$, i.e., $R_t = -|\Theta_t - r_t|$. If the system truncates ($|\Theta_t| \ge \pi/2$), the reward is scaled by the remaining time steps, penalizing premature truncation.
\end{itemize}

The test task is structured as follows: the system aims to achieve a sequence of target angles over an \SI{80}{\second} time frame with a sample time of \SI{0.1}{\second}.
Every \SI{10}{\second}, the target pitch is updated to a new value, following the predefined profile \ang{0}, \ang{5}, -\ang{5}, \ang{20}, -\ang{20}, \ang{40}, -\ang{40}, \ang{0}.
This formulation provides a controlled setting to evaluate the agent's ability to track a series of reference angles.

\section{Crucial Aspects of Problem Formulation} \label{sec:hypotheses}

Effective problem formulation is essential for successful \ac{rl} applications in \acp{icps}.
Subtle adjustments to states, actions, and rewards can significantly impact training efficiency and performance.

\subsection{Improvements to the Problem Formulation}
Building on the core components of \acp{mdp}, we propose five hypotheses (A--E) to enhance \ac{rl} outcomes in this context:
\begin{enumerate}[label=\Alph*)]
    \item \emph{Normalization Improves Convergence and Sample Efficiency:}
    Large or inconsistent input ranges can destabilize gradient updates in deep \ac{rl}. Normalization of states, actions, or rewards can help constrain the domain of the learning signal, improving gradient flow and accelerating exploration.

    \item \emph{Randomizing Target References Improves Generalization:}
    With a fixed target trajectory, the agent overfits rather than learning a more flexible control policy. Randomizing the target references across episodes broadens the range of scenarios encountered during training.

    \item \emph{Longer Episodes Provide Richer Trajectories and Accelerate Learning:}
    Short episodes can limit the agent's exposure to delayed rewards and reduce state-space exploration.

    \item \emph{Randomizing Initial States Encourages Better Exploration:}
    Always starting from the same initial condition restricts early exploration to a narrow region of the state space.


    \item \emph{Combining All Factors Leads to Stable and High-Performing Convergence:}
    Each of the above techniques addresses weaknesses in \ac{rl} problem formulation.
    When applied together, these techniques should create a complementary effect, leading to improved training stability, faster convergence, and higher overall performance by addressing multiple challenges simultaneously.
\end{enumerate}

To test these hypotheses, we define two main configurations: (i) the original problem definition from \cref{sec:original-problem-formulation} called \enquote{Baseline}, and (ii) applying all improvements simultaneously to a configuration called \enquote{New setting}. The differences of those two environments is summarized in \cref{tab:parameter-comparison}.

\begin{table}[ht]
\centering
\caption{Parameter comparison of \enquote{Baseline} and \enquote{New setting}.} \label{tab:parameter-comparison}
\begin{tabular}{@{}l|ll@{}} 
\toprule
               & \textbf{Baseline} & \textbf{New setting} \\ \midrule
stop time      & \SI{80}{\second}  & \SI{100000}{\second} \\
target tilt    & fixed             & random               \\
initial tilt   & \ang{0}             & random               \\
norm obs       & no                & yes                  \\
norm action    & no                & yes                  \\
action penalty & 0.0               & 0.25                 \\ \bottomrule
\end{tabular}
\end{table}

\subsection{Studies on the Influence of Suggested Improvements}
For each hypothesis A--D, we conduct an addition and an ablation study:

\begin{itemize}
    \item \emph{\enquote{Baseline} with \enquote{New} Parameter:}
    We start from the \enquote{Baseline} and apply one change from the \enquote{New setting} (e.g., just add normalization).
    \item \emph{\enquote{New Setting} with \enquote{Baseline} Parameter:}
    We start from the \enquote{New setting} and change that parameter to match the \enquote{Baseline} configuration (e.g., remove normalization).
\end{itemize}

This isolates the direct effect of each parameter.
We use \ac{ppo} from the \ac{sb3} library \cite{stable-baselines3} with default hyperparameters and train for 1 million steps, repeating each experiment over 10 seeds.
Every \num{10000} steps, an evaluation run is triggered.
This allows us to record (i) how many steps it takes to reach an average deviation of \ang{4} or better on the evaluation profile, (ii) final performance metrics (mean/min/max/standard deviation of pitch deviation) after 1 million steps, (iii) the average absolute voltage applied, indicating how aggressively the agent acts. Additionally, we conduct a training run on the real system using the parametrization from the \enquote{New setting}, starting the training process from scratch without utilizing any pretrained agent.

\subsection{Implementation Details of Suggested Improvements}

\subsubsection{Stop Time}
The simulation horizon was increased from \SI{80}{\second} to \SI{100000}{\second}, allowing the agent to explore a longer sequence of states.
With \num{1000000} training steps and a sample time of \SI{0.1}{\second} this results in a single episode per training run if no truncation occurs.

\subsubsection{Random Target Tilt}
Rather than training with a single, fixed target tilt profile, the \enquote{New setting} environment dynamically changes the target.
At each time step, there is a 1\% probability that the target tilt is randomly redrawn from the range of \SIrange{-40}{40}{\degree}.
On average, this results in a target change every \SI{10}{\second} (i.e., every \num{100} steps), ensuring that the agent does not converge to a narrow policy specialized to a single tilt profile.
Introducing this variation fulfills the Markov property, as the environment state no longer follows a static target.

\subsubsection{Random Initial Tilt}
When the environment is reset, the tilt angle is uniformly sampled from a range of \SIrange{-40}{40}{\degree} in the \enquote{New setting}, instead of being fixed at \ang{0}.
This introduces additional variability into each trial and forces the agent to adapt from a variety of starting positions.

\subsubsection{Observation Normalization}
In the \enquote{New setting}, each component of the observation vector is normalized by its maximum expected magnitude.
Specifically, the pitch and target angles are divided by $\pi/2$, while the velocity is scaled by a factor of $0.2441$, based on empirical measurements.
This normalization ensures that the inputs to the agent remain within specific ranges (\SIrange{-1}{1}{}) and helps stabilize the learning process.

\subsubsection{Action Normalization}
The action space is rescaled from direct voltages values from the range of \SIrange{-24}{24}{\volt} to a normalized range of \SIrange{-1}{1}{}.
To apply the action in the actual system, the agent's chosen action $a \in [-1, 1]$ is multiplied by the maximum voltage of \SI{24}{\volt}.
Hence, the true control voltage becomes $\text{voltage} = 24 \cdot a$.

\subsubsection{Action Penalty}
Finally, an action penalty factor was added to the reward function in the \enquote{New setting} to discourage large fluctuations in the applied voltage.
Over a 1-second window, the standard deviation of the normalized applied voltage is computed and then multiplied by the action penalty factor (\num{0.25}).
This term is subtracted from the original reward, thus encouraging smoother voltage profiles.

By incorporating these modifications, either individually (for the hypothesis tests A--D) or collectively (in the full \enquote{New setting}) for hypotheses E, we can isolate and examine the effect of each parameter choice and observe how it impacts the agent's learning progress.

\subsection{Remark on Real-World Applications}
When training \ac{rl} on a physical setup, additional considerations are necessary to preserve hardware integrity and ensure safe operation.
In simulation, aggressive of \enquote{bang-bang} control policies may appear highly effective, as virtual environments do not exhibit real-world wear-and-tear.
In contrast, physical motors, actuators, and sensors face constraints such as temperature limits, friction, and mechanical stress.
Repeatedly applying large, abrupt control signals can lead to overhearing, premature failures, or damage to the system's components.
To mitigate these risks, we introduce an action penalty term that motivates smoother, more gradual voltage changes.
This adjustment not only improves the longevity of the hardware but also encourages the agent to learn control policies that avoid abrupt power surges.

\section{Experimental Results and Discussion}
In \cref{tab:experiment-comparison}, we summarize the performance of the different experiments based on the results of 10 repeated training runs each. For each metric and each experiment we assume that the outcomes follow a normal distribution. To compare two experiments for a given metric, we essentially ask for the probability, when we draw a sample of each, that the value for one experiment would be less (or greater) than the other.


Say we have $X_1 = \mathcal{N}(\mu_1, \sigma_1)$ for experiment 1 and $X_2 = \mathcal{N}(\mu_2, \sigma_2)$ for experiment 2 then $X_1 - X_2$ is distributed $\mathcal{N}(\mu_1 - \mu_2, \sqrt{\sigma_1^2 + \sigma_2^2)}$ and we ask for the probability $P(x_1 < x_2)$ when $x_1 \sim X_1, x_2 \sim X_2$. This leads to the score
\[
z(X_1, X_2) = \frac{\mu_{1} - \mu_{2}}{\sqrt{{\sigma_1}^2 + {\sigma_2}^2}}
\]
used as follows: $P(x_1 < x_2)$ is equal to the probability that $N(0, 1)$ gives a value greater than $z(X_1, X_2)$, i.e., the smaller $z(X_1, X_2)$ the higher the probability for $x_1 < x_2$. This underpins the following notation:
We call $X_1$ is significantly better (smaller) than $X_2$ (indicated by \textcolor{improvement}{$\uparrow$}) when $z(X_1, X_2) < -1$; we call $X_1$ significantly worse (greater) than $X_2$ (indicated by \textcolor{decline}{$\downarrow$}) when $z(X_1, X_2) > 1$, and otherwise there is no significant difference (indicated by $\sim$).

\subsection{Analysis of Hypotheses (A--E)}

We summarize our findings for each hypothesis below:

\begin{itemize}
    \item \emph{Normalization (Hypothesis A):}
    Normalizing states and actions emerges as a key factor in stabilizing and expediting training.
    In the \enquote{Baseline} setting, normalization alone enables the agent to reach the \ang{4} threshold in every run.
    In the \enquote{New setting}, removing normalization significantly degrades performance.
    Additional experiments isolating action and state normalization confirm that action normalization exerts a greater influence in our case: the observation space already lies in a relatively low range, while action space spans \SIrange{-24}{24}{\volt}.
    
    \item \emph{Random Targets (Hypothesis B):}
    Allowing the agent to experience a variety of target reference angles during training improves sample efficiency in the \enquote{Baseline} configuration and does not yield a significant impact by removing it from the \enquote{New setting}.
    To analyze the actual impact of the generalizability, additional evaluations runs must be conducted using a variety of target trajectories, which is out of scope of this work.
    
    \item \emph{Longer Episodes (Hypothesis C):}
    Increasing the horizon provides more continuous state-space exploration before resets occur.
    In the \enquote{Baseline}, this markedly enhances performance, whereas in the \enquote{New setting} its effect is less pronounced.
    
    \item \emph{Random Initial Pitch (Hypothesis D):}
    Introducing a stochastic range of initial pitch angles encourages better coverage of the state space during early training.
    Although it greatly benefits the \enquote{Baseline}, in the \enquote{New setting} the effect is smaller, as long episodes reduce reliance on any single initial condition.
    
    \item \emph{Combining All Factors (Hypothesis E):}
    Aggregating these individual improvements yields the fastest and most stable convergence.
    The \enquote{New setting} outperforms the \enquote{Baseline} in both sample efficiency and final performance, underscoring the complementary nature of these design choices.
\end{itemize}

Even though the \enquote{Baseline} configuration appears to yield lower action magnitudes, this can be deceptive: the agent may simply fail to achieve the goal and therefore minimizing control effort (i.e., doing almost nothing).
By contrast, the \enquote{New setting} applies notably high control signals, effectively resembling a \enquote{bang-bang} strategy.
Introducing an action penalty term into the \enquote{New setting} does not degrade performance; in fact, it marginally improves the final policy while substantially reducing abrupt control actions, thereby mitigating hardware stress when deployed to the real system.

\begin{table*}[t]
\centering
\caption{Comparison of the different experiments derived from the Hypotheses (A--E).} \label{tab:experiment-comparison}
\begin{tabular}{@{}l|ll|ll|ll@{}}
\toprule
\textbf{Method} & \textbf{steps to \ang{4}} & $\boldsymbol{z}$ & \textbf{deviation (\unit{\degree})} & $\boldsymbol{z}$ & \textbf{voltage (V)} & $\boldsymbol{z}$ \\ \midrule

Baseline & n/a $\pm$ n/a (\SI{0}{\percent}) &  & 6.88 $\pm$ 1.94 &  & 2.81 $\pm$ 0.37 &  \\ \midrule \midrule

Baseline with normalization & \num{63000} $\pm$ \num{22825} (\SI{100}{\percent}) & n/a \improve & 2.91 $\pm$ 0.23 & -2.03 \improve & 8.78 $\pm$ 5.42 & 1.10 \decline \\ 

Baseline with random targets & \num{940000} $\pm$ \num{0} (\SI{10}{\percent}) & n/a \improve & 6.37 $\pm$ 1.66 & -0.20 \neutral & 2.84 $\pm$ 0.38 & 0.06 \neutral \\ 

Baseline with long episodes & \num{886667} $\pm$ \num{26247} (\SI{30}{\percent}) & n/a \improve & 5.47 $\pm$ 2.04 & -0.50 \neutral & 3.15 $\pm$ 0.79 & 0.39 \neutral \\ 

Baseline with random initial pitch & \num{872500} $\pm$ \num{83179} (\SI{40}{\percent}) & n/a \improve & 4.80 $\pm$ 1.21 & -0.91 \neutral & 3.37 $\pm$ 0.55 & 0.85 \neutral \\ \midrule \midrule

New setting & \num{40000} $\pm$ \num{11832} (\SI{100}{\percent}) &  & 3.05 $\pm$ 0.15 &  & 21.47 $\pm$ 2.28 &  \\ \midrule

New setting without normalization & \num{595000} $\pm$ \num{123119} (\SI{60}{\percent}) & 4.49 \decline & 3.72 $\pm$ 1.27 & 0.53 \neutral & 3.84 $\pm$ 0.63 & -7.45 \improve \\ 

New setting without random targets & \num{42000} $\pm$ \num{22271} (\SI{100}{\percent}) & 0.08 \neutral & 2.88 $\pm$ 0.23 & -0.63 \neutral & 10.29 $\pm$ 6.36 & -1.65 \improve \\ 

New setting without long episodes & \num{35000} $\pm$ \num{9220} (\SI{100}{\percent}) & -0.33 \neutral & 2.97 $\pm$ 0.31 & -0.24 \neutral & 13.00 $\pm$ 6.12 & -1.30 \improve \\ 

New setting without random initial pitch & \num{34000} $\pm$ \num{14967} (\SI{100}{\percent}) & -0.31 \neutral & 3.05 $\pm$ 0.16 & 0.01 \neutral & 21.30 $\pm$ 2.22 & -0.05 \neutral \\ \midrule \midrule

New setting with action penalty & \num{26000} $\pm$ \num{9165} (\SI{100}{\percent}) & -0.94 \neutral & 3.08 $\pm$ 0.23 & 0.11 \neutral & 4.29 $\pm$ 0.19 & -7.51 \improve \\ \midrule

New setting with action penalty on real system & \num{200000} & & 3.58 & & 3.85 \\ \midrule

\end{tabular}
\end{table*}

\begin{figure}[ht]
    \centering
    \includegraphics[width=\linewidth]{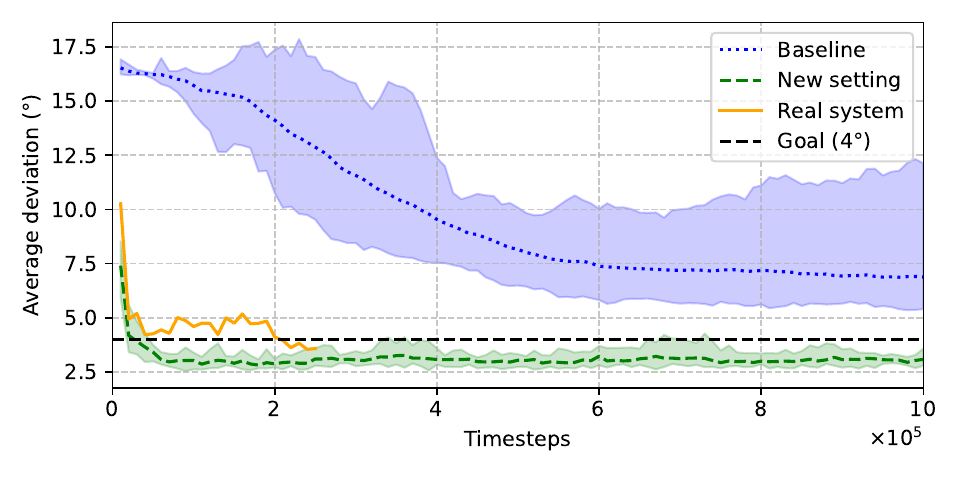}
    \caption{Average deviation to the target for the \enquote{Baseline} and \enquote{New setting} configurations on the evaluation profile during training.
    The \enquote{Baseline} and \enquote{New setting} configurations were trained on the simulation model for 1 million steps.
    Additionally, the plot includes the results for the \enquote{New setting with action penalty} configuration trained on the real system for \num{250000} steps.}
    \label{fig:training-model}
\end{figure}

\begin{figure}[ht]
    \centering
    \includegraphics[width=\linewidth]{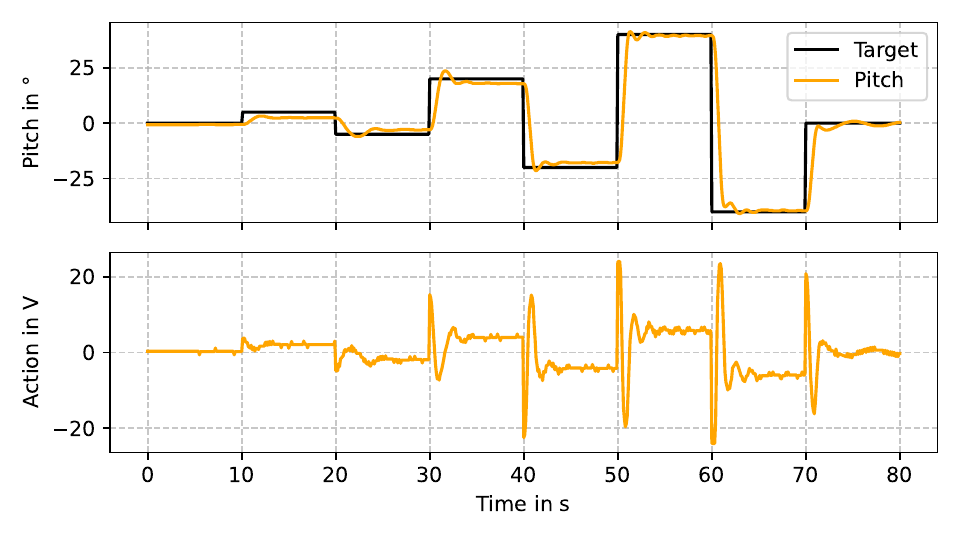}
    \caption{Evaluation on the real system with an agent trained for \num{250000} steps, showing the target pitch, actual pitch, and applied voltage.}
    \label{fig:evaluation-run}
\end{figure}

\Cref{fig:training-model} illustrates the training performance, measured as the average deviation from the target, over 1 million training steps.
The \enquote{New setting} demonstrates superior performance compared to the \enquote{Baseline}, excelling in sample efficiency, overall performance, and stability.
Training on the real system reached the predefined goal of \ang{4} within \num{200000} steps and quickly converged to within \ang{5} degrees.
\Cref{fig:evaluation-run} provides a detailed view of the actual pitch, the target, and the applied voltages of the agent trained on the real system after \num{250000} steps.

\section{Conclusion and Future Work}
This paper showed that minor yet carefully chosen adjustments to \ac{rl} problem formulation can significantly boost performance, training stability, and efficiency in \acp{icps}.
Our experiments on a \ac{1dof} helicopter testbed confirm that normalization, randomizing targets and initial states, extending episodes to horizons, and reward shaping all foster more reliable and sample-efficient learning.
Moreover, training directly on the real hardware without a prior model illustrates the feasibility of real-world \ac{rl} applications.
Future research will focus on establishing a robust \ac{rl} engineering pipeline for \acp{icps}, integrating advanced approaches such as data-driven probabilistic \ac{mpc} or physics-informed \ac{rl}, and incorporating additional performance metrics for further optimization.

\bibliographystyle{IEEEtran}
\bibliography{jrcisia-published, references}

\end{document}